\documentclass[letterpaper]{article}

\usepackage{aaai25}  
\usepackage{times}  
\usepackage{helvet}  
\usepackage{courier}  
\usepackage[hyphens]{url}  
\usepackage{graphicx} 
\usepackage[sort]{natbib}  
\usepackage{booktabs}
\usepackage{makecell}
\usepackage{tikz}
\usetikzlibrary{positioning, shapes, arrows, shadows, arrows.meta, fit}
\usepackage{caption} 
\usepackage{algorithm}
\usepackage{algorithmic}

\title{Niche Dynamics in Complex Online Community Ecosystems}
\author {
    Nathan TeBlunthuis
}
\affiliations{
    University of Texas at Austin\\
    School of Information\\
    nathante@utexas.edu
}

\begin{document}









\maketitle

\begin{abstract} 

Online communities are important organizational forms whe\-re
members socialize and share information. Curiously, different online
communities often overlap considerably in topic and membership. Recent
research has investigated competition and mutualism among overlapping
online communities through the lens of organizational ecology;
however, it has not accounted for how the nonlinear dynamics of online
attention may lead to episodic competition and mutualism. Neither has
it explored the origins of competition and mutualism in the processes
by which online communities select or adapt to their niches.  This
paper presents a large-scale study of 8,806 Reddit communities
belonging to 1,919 clusters of high user overlap over a 5-year
period. The method uses nonlinear time series methods to infer bursty,
often short-lived ecological dynamics.  Results reveal that mutualism
episodes are longer lived and slightly more frequent than competition
episodes. Next, it tests whether online communities find their niches
by specializing to avoid competition using panel regression models.
It finds that competitive ecological interactions lead to decreasing
topic and user overlaps; however, changes that decrease such niche
overlaps do not lead to mutualism. The discussion considers future
designs for online community ecosystem management. \end{abstract}

\section{Introduction}

Online communities are increasingly important sites where
people organize to meet their members' various needs to
produce and consume information goods, provide social support
\cite{de_choudhury_mental_2014}, engage in collective and
political action \cite{choudhury_social_2016, benkler_social_2013,
krafft_disinformation_2020}, make sense of the world, and connect
with each other \cite{benkler_wealth_2006, lampe_motivations_2010}.
Individuals often belong to overlapping online communities that
have suprisingly similar members and topics \cite{teblunthuis_no_2022,
datta_identifying_2017, tan_all_2015}. Why are there so many
overlapping online communities? How do they come to have the
topics that they do? Why do overlapping sets of users so often
participate in communities about superficially similar topics
\cite{teblunthuis_no_2022, datta_identifying_2017}?

Interdisciplinary scholarship has explored these questions through the
lens of organizational ecology \cite{teblunthuis_identifying_2022,
zhu_impact_2014, zhu_selecting_2014, wang_impact_2012}.
Organizational ecology considers the role of environmental
and relational forces in the development of organizations
and industries \cite{hannan_organizational_1989,
mcpherson_ecology_1983, hawley_human_1986, aldrich_organizations_2006,
van_de_ven_explaining_1995}. Early studies in this
vein used density dependence theory to relate the overlaps
in community membership to growth and survival
\cite{zhu_impact_2014,zhu_selecting_2014,wang_impact_2012}.  Recently,
\citet{teblunthuis_identifying_2022} introduced an alternative
approach using time-series analysis to directly infer competitive
and mutualistic relationships; however, their analysis is based on an
unrealistic assumption in the online community context. They
modeled intercommunity ecological relationships as static over
time. However, the dynamics of attention online are often bursty
\cite{ratkiewicz_characterizing_2010}.

This study addresses this limitation using nonlinear time-series
methods to infer competitive and mutualistic interactions between
online communities that can vary over time. Specifically, it uses the S-Map model developed by ecologists of biological systems for this purpose \cite{cenci_regularized_2019,sugihara_nonlinear_1994,sugihara_nonlinear_1990}.
\citet{teblunthuis_identifying_2022} found that
mutualism was much more common than competition; other
ecological studies of online communities have similarly pointed to
specialization and mutualism as crucial for online community
success \cite{zhu_impact_2014, teblunthuis_no_2022}. Therefore, this study tests the hypothesis that \emph{mutualism will be more common than competitive}.

This study also tests a model inspired by theories of specialization
and adaptation in organizational ecology to explain how online
communities find their \emph{niches} and how highly overlapping
online communities develop \cite{dobrev_evolution_2002,
baum_ecological_2006}.  Competition and mutualism might result from
niche overlap, but might also cause communities to shift their niches.
This stu\-dy investigates such feedback between niche and ecological
dynamics by combining the S-Map's measures of ecological interactions
with longitudinal measures of user and topic similarity in a panel
data analysis of 48,484 relationships between
8,806 online communities on Reddit over a 5 year time span.  If online communities adapt in response
to competition, \emph{then more competitive ecological interactions
should predict decreases in overlap}. If specialization helps
communities avoid competition, then \emph{decreases in overlap should
predict increases in mutualism}.


This work contributes to the social science of online communities by
(1) inferring time-varying nonlinear ecological dynamics, (2) shedding
new light on findings from prior work, (3) engaging the organizational
ecology concepts of structural intertia and competitive exclusion
to theorize the role of adaptation in producing the widespread
mutualism observed in prior studies and, (4) testing the resulting
hypotheses to (5) synthesize insights for leaders and designers of
online communities. The nonlinear time-series analysis reveals that
ecological relationships tend to happen in bursty episodes and that
mutualism has similar strength to competition; however mutualism
happens more often and for longer durations. The panel models of
niche adaptation found that online communities in competition tend
to reduce their overlaps; however such changes lead to increases,
not decreases in competition. The latter result is consistent
with previous research findings that online communities tend to
become oligarchical and rigid \cite{teblunthuis_revisiting_2018,
hill_studying_2019}, and that organizational change often has
negative outcomes \cite{hannan_organizational_1989}. Online community
designers and leaders can attempt to manage online ecosystems by
filling ecological roles.  Feed or recommendation algorithms can be
designed to adapt existing communities' niches by purposfully nudging
individuals. That said, given the challenges of organizational change,
a more pragmatic approach is likely to building new communities as
spin-offs or offshoots from existing ones.

\section{Background}

\subsection{Interactions Between Changing Online Communities}

Online communities are a dynamic, growing, and increasingly important form of social organization. Peer production communities such as Wikipedia and open-source software projects have produced tens of billions of dollars' worth of software and encyclopedic documentation \cite{benkler_peer_2015}. Millions of members of smaller communities on platforms such as Reddit, Facebook, and Discord organize political mobilizations including disinformation campaigns and protest movements \cite{choudhury_social_2016, benkler_social_2013, krafft_disinformation_2020} and exchange social support \cite{de_choudhury_mental_2014}, entertainment, and information \cite{benkler_wealth_2006}. Online community platforms support millions of attempts to build communities \cite{hill_studying_2019, schweik_internet_2012}; however, only a tiny percentage manages to mobilize participants and sustain collaboration \cite{schweik_internet_2012}.  

Early research into the growth, survival, and success of online communities focused almost exclusively on the perspective of managers of a single community to find effective practices for motivating and sustaining productive participation \cite{kraut_building_2012}.
Explanatory variables included the characteristics of founders \cite{kraut_role_2014, kairam_how_2024}, language use \cite{danescu-niculescu-mizil_no_2013}, turnover \cite{dabbish_fresh_2012}, and designs for regulating behavior \cite{halfaker_rise_2013, teblunthuis_revisiting_2018}.
Analyses from this ``focal organization perspective'' \cite{hannan_organizational_1989} have typically accounted for only a small amount of variation in communities' growth, longevity, and performance and have limited ability to predict community growth or long-term participation \cite{cunha_are_2019}.

Recent scholarship has demonstrated that interdependence among
online communities is widespread, important to explaining success
and failure, and likely to provide new insights for designing and
managing communities \cite{cunha_are_2019, mitts_removal_2022,
chandrasekharan_you_2017, kairam_life_2012, tan_all_2015,
tan_tracing_2018, teblunthuis_no_2022, vincent_examining_2018}.  Most
pertinent are studies that adopt the theoretical lens
of \emph{organizational ecology} to understand \emph{competition}
and \emph{mutualism} between overlapping online communities
\cite{teblunthuis_identifying_2022, wang_impact_2012, zhu_impact_2014,
zhu_selecting_2014}.  The initial work in this line emphasizes
competition and finds that overlapping Usenet groups are likely
to compete \cite{wang_impact_2012}. Similarly, in chapter 6 of the
prominent textbook \emph{Building Successful Online Communities},
\citet{resnick_starting_2012} recommend that new online communities
``carve out a useful and defendable niche in the ecology of competing
communities.''

However, most empirical findings in ecological studies of online communities point to \emph{mutualism} and \emph{specialization} as beneficial to the success of overlapping communities \cite{teblunthuis_identifying_2022, zhu_impact_2014, zhu_selecting_2014, teblunthuis_no_2022}. 
Recent interviews with members of overlapping subreddits revealed that each community often provided benefits that others did not such as audience, belonging, and specific content and information \cite{teblunthuis_no_2022}.
Studying overlapping fandom Wikis, \citet{zhu_impact_2014} suggest that although communities might ``compete over shared members' time and efforts,'' overlapping communities share knowledge, diverse perspectives, and opportunities to recruit new members. Although the researchers hypothesized that competitive and mutualistic forces would trade off as user overlap increases, they instead found that increasing overlap was associated with increasing survival of new communities. A related study found evidence of both competition and mutualism; however mutualism was stronger than competition for most communities \cite{zhu_selecting_2014}. \citeauthor{teblunthuis_identifying_2022} inferred networks of competition and mutualism among overlapping subreddits using time-series analysis and found that mutualism is more common than competition \cite{teblunthuis_identifying_2022}. 

Yet the method \citeauthor{teblunthuis_identifying_2022} propose for
inferring competition and mutualism depends on vector autoregression
(VAR) models.  Although these models have been used in ecology
to infer competition and mutualism \cite{ives_estimating_2003},
they also bear assumptions that may be unrealistic in online
communities. VAR models can only represent linear dynamics and
competitive and mutualistic interactions that do not vary over
time \cite{cenci_non-parametric_2019, cenci_regularized_2019};
however, online communities inhabit dynamic environments and
experience shocks such as influx of newcomers and attention
\cite{zhang_participation_2019, kiene_surviving_2016, lin_better_2017,
ratkiewicz_characterizing_2010}. For example, policy changes and bans
can influence related communities \cite{chandrasekharan_you_2017,
ribeiro_platform_2021, matias_going_2016}.  Therefore, this study uses
nonlinear time-series analysis to investigate how ecological relationships
in clusters of overlapping communities vary over time.  The first
hypothesis tests whether the linearity assumptions would not invalidate
previous findings by \citeauthor{teblunthuis_identifying_2022} and that
these reflect the average relationships over time.  H1: \emph{Mutualistic
interactions will be more frequent and longer lasting than competitive
interactions.}

\subsection{Explaining Widespread Mutualism}

How do systems of overlapping mutualistic online communities come
to be? How do these communities avoid competing with each other?
Measuring time-varying ecological interactions opens a path to test whether
adaptation can explain why mutualism appears more common than
competition.  Prior work drawing on organizational ecology have suggested
that online communities depend on types of resources with the potential
to create both mutualism and competition \cite{wang_impact_2012,
zhu_selecting_2014, teblunthuis_identifying_2022}.  Online communities
might compete for participants' time and attention, which is
a rival resource in that its usefulness decreases when it
is used. On the other hand, overlapping online communities can be
mutualists as they produce non-rival or ``anti-rival'' goods,
such as information, that increase in value with use and that other
communities can utilize. If resources are the whole story, then
widespread mutualism suggests that the non-rival resources shared
among communities are more important, a view resonant with early
optimism about online collaboration \cite{benkler_wealth_2006}.
However, an explanation based soley on resources does not account
for how online community builders are knowledgeable agents with specific
goals. Online communities do not spring from the ether; they are built
by their leaders and participants.

This section proposes that online communities' participants
construct systems of mutualistic communities that provide a
range of complementary benefits as they collectively adapt
existing communities to fill available niches. The competitive
exclusion principle---one of the most important concepts in ecology
\cite{armstrong_competitive_1980, hardin_competitive_1960}---proposes
that natural selection disfavors competitive relationships. Strong
enough competition between two groups will lead to the death of
at least one of the groups. Coexistence is possible only through
specialization \cite{levin_community_1970}.  Recall recent qualitative
work that finds that overlapping communities provide different types
of benefits to their users \cite{teblunthuis_no_2022}. Ecological
theory suggests that such specialization will reduce competition.

A two-stage process may occur in which competitive dynamics first
lead to specialization, and this specialization subsequently
decreases competition.  Classical organizations facing competition
may attempt to change and differentiate from competitors
\cite{mcpherson_evolution_1991}.  However, organizations often
fail to affect such change \cite{hannan_structural_1984} because
forces of institutional isomorphism \cite{dimaggio_iron_1983},
contagion \cite{greve_jumping_1995}, or organizational
learning \cite{dobrev_shifting_2003} may be more important
causes of organizational change than competition avoidance.
Furthermore, forces termed ``structural inertia'' also hinder
change efforts, including organizational culture, internal
patronage networks, conflicts among stakeholders, and routines
\cite{hannan_structural_1984, van_de_ven_explaining_1995}. Such
inertial forces can also increase failure rates following periods of
change \cite{hannan_organizational_1989}.

Although structural inertia is central to organizational eco\-logy, prior ecological studies of online communities have barely touched on it.
However, empirical findings strongly suggest that online communities also experience structural inertia.
Online communities tend to change their policies less frequently over time \cite{teblunthuis_revisiting_2018, halfaker_rise_2013} and have little turnover in their leaders who often resist change \cite{shaw_laboratories_2014}.
Similar to classical organizations, online communities have roles \cite{arazy_how_2017, arazy_functional_2015}, routines \cite{keegan_analyzing_2016} and concertive control \cite{gibbs_digital_2021}, all of which can lead to structural inertia.
Moreover, online communities typically lack capacities to coerce or incentivize change available to classical organizations such as firms and governments. Thus, endogenous structural changes in established online communities probably occur in bottom-up processes driven by the choices of individual members or small groups of them \cite{steinsson_rule_2023}.

Even without capacities for top-down structural change, members may tend to  contribute to communities in ways that lead to specialization \cite{van_koevering_whats_2024}. 
Online communities are typically ``open'' organizations where individuals can freely participate in multiple communities at once and share or repost the same content in different communities \cite{butler_cross-purposes_2011}.  
When some members become unhappy, an online community characterized by structural inertia is unlikely to change policies or leadership to satisfy them.
However, in a system of overlapping open communities, unhappy members can ``exit'' and migrate to an overlapping community and bring their contributions along \cite{frey_emergence_2018, hirschman_exit_1970}.

 In ecological terms, such migrations correspond a competition episode between the communities because a decrease in participation in one group coincides with an increase in participation in the other.
Migration may also decrease the overlap of the two communities' niches, as recent empirical work on Reddit suggests \cite{van_koevering_whats_2024}.
 If individuals participate more in communities where they have the greatest expectation of finding a specific type of benefit,
their participation can reinforce the community's ability to provide these specific benefits. Their contributions can increase the supply of content, attention, and social interaction, and they may reward others who make similar contributions with thanks, awards, votes and other signals of approval. 
The size of the group of individuals desiring these specific benefits may subsequently increase in the second community and decrease in the first.
The aggregated actions of individuals with aligned interests can collectively result in a positive feedback loop that increases and stabilizes the second community's provision of these benefits while decreasing the niche overlap between the two communities in the process \cite{mcpherson_testing_1996, schelling_micromotives_1978}. 

Previous ecological studies of online communities have measured niches by quantifying user and topic overlaps \cite{zhu_impact_2014, teblunthuis_identifying_2022,wang_impact_2012,zhu_selecting_2014}. Logically, specialization might happen along either dimension or both. Therefore,  the following two hypotheses test whether online communities specialize by reducing niche overlap.

H2a: \emph{Two subreddits having greater competition (mutualism) will subsequently have greater decreases (increases) in \textbf{topic overlap}.}

H2b: \emph{Two subreddits having greater competition (mutualism) will subsequently have greater decreases (increases) in \textbf{user overlap}.}

The competitive exclusion principle predicts that specialization reduces competition over rival resources \cite{armstrong_competitive_1980, hardin_competitive_1960}. Resource partitioning theory and niche width theory in organizational ecology explain the relationship between competition and overlap \cite{dobrev_shifting_2003, dobrev_dynamics_2001, carroll_concentration_1985}. Sometimes, organizations can adapt to avoid competition as when automotive manufacturers with greater niche overlaps undergo greater transformations during periods of technological change \cite{dobrev_shifting_2003}. Such change can be risky; however increasing specialization to avoid competition can be necessary for organizational survival \cite{baum_ecological_2006}. 
Importantly, the ecological interactions between two online communities can involve a mixture of mutualistic and competitive components.  For instance, two communities may compete over contributors' time and effort while simultaneously benefiting from sharing information or content.
Even if the relationship between two communities is mutualistic overall, a decrease in niche overlap can reduce the competitive component of the relationship and increase the overall mutualism.  Therefore, the final two hypotheses test whether increases (decreases) in niche overlaps will lead to subsequent (decreases) increases in (mutualism) competition.  

H3a: \emph{Two subreddits having decreasing (increasing) \textbf{topic overlap} will subsequently have greater mutualism (competition).}

H3b: \emph{Two subreddits having decreasing (increasing) \textbf{user overlap} will subsequently have greater mutualism (competition).}

\section{Data and Measures}
\subsection{Online Communities Hosted on Reddit.com}
This study requires a dataset of communities that are all active
over a sufficiently long period; it uses the publicly available
Pushshift archive of Reddit submissions and comments with a
study period of December 5\textsuperscript{th} 2015 to April
13\textsuperscript{th} 2020 \cite{baumgartner_pushshift_2020}.
The included subreddits had comments or submissions in at least 20\% of
weeks during the study period; however, they did not have more than 10\% of
submissions marked ``not safe for work'', which often indicates sexual
content.

\subsection{Overview} 
\begin{figure*}
 \includegraphics[width=\linewidth]{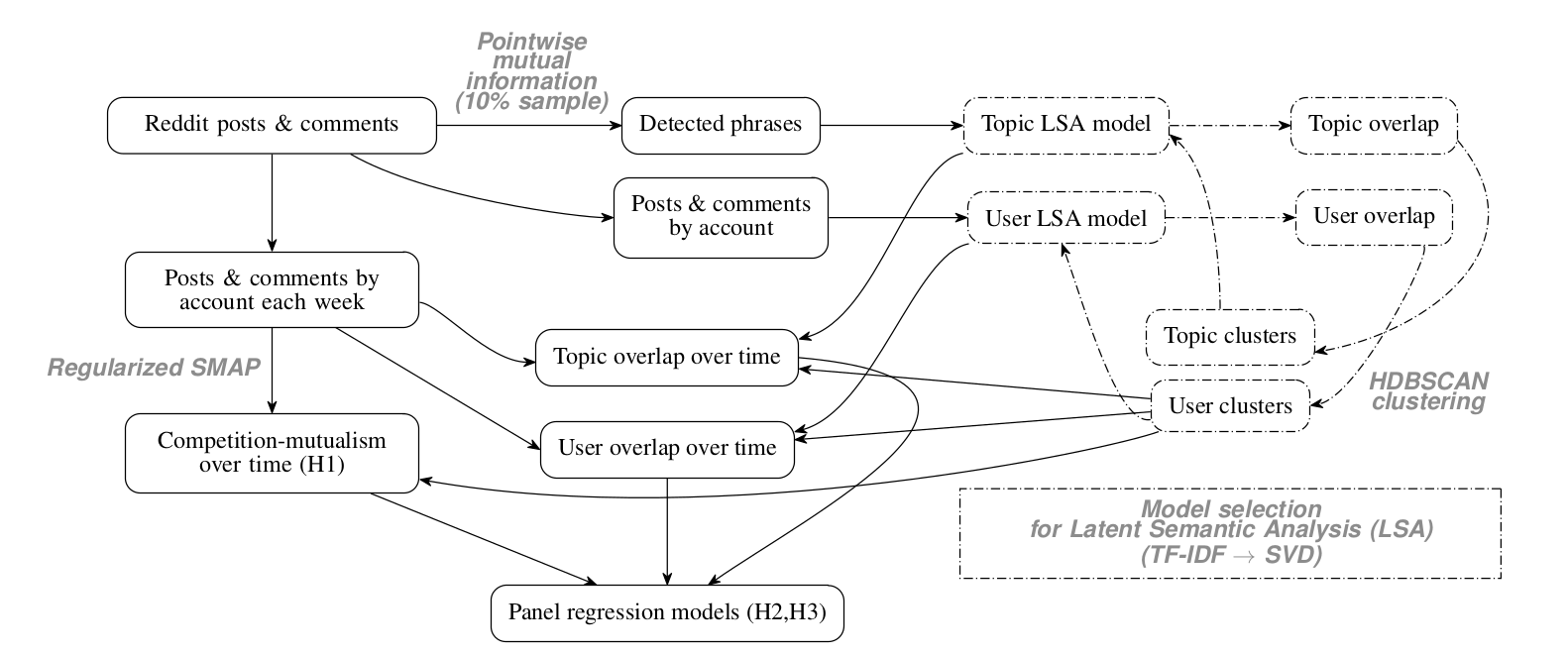}

\caption{Flowchart representing the pipeline for building longitudinal measures of competition and mutualism as well as topic and author overlap and testing the hypotheses. Hyperparemters for LSA models are selected in a repeated process to find a good silhouette coefficient for the topic and user clusters. \label{fig:methods.flowchart}}
\end{figure*}

The analysis proceeds as follows and summarized in Figure \ref{fig:methods.flowchart}: First, a
clustering pipeline discovers groups of highly overlapping subreddits
based on membership similarity. The S-Map algorithm then measures
competition and mutualism episodes with which to test H1. Finally,
the panel regression models test H2 and H3 using weekly similarity
measures based on membership and content and competition and
mutualism that the S-Map measured. The next section documents the
measures and analytic plan, and the results follow.


\subsection{Discovering Subreddit Clusters} 

\label{sec:mes.overlap}

A procedure similar to that of \cite{teblunthuis_identifying_2022}
constructs clusters of overlapping subreddits with overlapping
users. The first step in this process is to measure \emph{user
overlap} $o_{i,j}$ between each pair of subreddits $i$ and $j$. This
is the number of contributions (posts or comments) made by each user
account in each subreddit normalized by the maximum number of
contributions to the subreddit by any account. Next, Latent Semantic
Analysis (LSA) with 2,000 dimensions reduces the dimensionality of the
frequency vectors. 2,000 dimensions for LSA gave the best clustering
performance as described in the next paragraph. The cosine similarity
of the resulting vectors for each pair of subreddits computes user
overlap.

The HDBSCAN algorithm clusters subreddits into groups
with high user overlap over the entire study period.
\citeauthor{teblunthuis_identifying_2022} tested different
clustering algorithms and found that HDBSCAN worked the best in terms
of the silhouette coefficient \cite{rousseeuw_silhouettes_1987}, a
measure of clustering performance that quantifies within-cluster
similarity. A grid sea\-rch finds HDBSCAN hyperparameters and the number
of LSA dimensions that together yeild a high silhouette coefficient as long
as fewer than 10,000 subreddits are not assigned to any cluster
because removing subreddits increases the silhouette coefficient.  The
chosen solution had a silhouette coefficient of (0.53), 1,949
clusters and 9,833 isolated subreddits.

\subsection{Inferring Dynamic Ecological Interactions}

Time-series models of \emph{group size} infer ecological
interactions. Group size is the number of distinct contributors
to the subreddit each week.  As the number of groups increases,
our method requires an exponentially longer time series
\cite{cenci_regularized_2019}; thus 30 clusters
that have more 28 communities were excluded.  The final
dataset has 8,806 subreddits in
1,919 clusters with 48,484
relationships measured 17,374,116 times over up to
758 weeks.  Table \ref{tab:sub.stats} summarizes the
activity levels in these communities.

\begin{table}

\centering
\begin{tabular}[t]{lccccc}
\toprule
\textbf{\makecell[l]{Weekly \\ Messages}} & \textbf{Min} & \textbf{Mean} & \textbf{Med} & \textbf{Max} & \textbf{Std. Dev}\\
\midrule
Min & 0.00 & 92 & 0 & 47548 & 858\\
Mean & 0.43 & 378 & 40 & 78451 & 2032\\
Median & 0.00 & 323 & 26 & 78090 & 1890\\
Maximum & 3.00 & 1516 & 201 & 325882 & 7224\\
Std. Dev & 0.45 & 238 & 35 & 64956 & 1191\\
\bottomrule
\end{tabular}
\caption{\label{tab:sub.stats}Statistics summarizing the distribution of contributions made in subreddits. The rows are statistics within each subreddit during the study period. The columns are statistics over the subreddits.}
\end{table}

As discussed above, adaptation to changing environmental
conditions or exogenous shocks can drive changes in the interactions between
online communities over time. This stu\-dy measures
these dynamic competitive and mutualistic interactions
between online communities using the regularized S-Map, a
nonlinear, nonparametric model that uses time-series data
of population sizes to infer an evolving matrix of community
interactions \cite{cenci_regularized_2019, deyle_tracking_2016,
cenci_assessing_2020, cenci_non-parametric_2019}. Intuitively, this
model asks ``when the system has been in a state similar to the
present, what usually happens next?'' and uses the answer to generate
forecasts.

The S-map estimates a sequence of Jacobian matrices $\mathcal{C}^t$
quantifying community interactions at each time $t$; it models each
observation as a linear regression of past observations, weighted
by their similarity to the present. The weights are defined by a
distance function and a scaling kernel.  Following common choices in
EDM research and S-Map applications, the distance function is the
Euclidean norm, and the scaling kernel is the exponential kernel.

This kernel has a hyperparameter $\theta$ controlling the degree of
locality (closeness in states of the system) in the weights. With
$\theta \rightarrow 0$, the S-Map becomes a linear VAR model. Larger
values of $\theta$ correspond to greater degrees of nonlinearity
\cite{sugihara_nonlinear_1994}.

The S-map contains parameters for each pair of communities at
each time point and thus is prone to overfitting. Therefore,
elastic net regularization is used to improve numerical stability
and reduce the risk of incorrectly interpreting the noise introduced
by endogenous or unobserved factors as community interactions
\cite{cenci_regularized_2019}.

For each cluster, the procedure given by
\citeauthor{cenci_non-parametric_2019} selects a
model. A grid search based on leave-one-out cross validation
\cite{cenci_non-parametric_2019} finds the hyperparameters $\theta$
(locality), $\alpha$ (ratio of $l1$ to $l2$ penalization), and
$\lambda$ (total regularization penalty). Following previous applications
of regularized S-map models, data on group sizes are log-transformed
and standardized.

\subsection{Mutualistic and Competitive Episodes}

This study tested H1 by analyzing the prevalence and longevity
of competitive and mutualistic interactions.  For each tuple
of communities $(a,b)$ in a cluster, \emph{episodes} of
competition or mutualism are consecutive weeks when the element
of the Jacobian $\mathcal{C}_{a,b}^t$ is positive or negative.
Ecological interactions may be asymmetrical; therefore, for
each pair of communities ${a,b}$, episodes of interaction
from $a$ to $b$ ($\mathcal{C}_{a,b}^t$) may differ from $b$
to $a$ ($\mathcal{C}_{b,a}^t$). The study includes a total of
17,374,116 episodes.

\subsection{Measuring Subreddit Similarity Over Time}

Measuring \emph{weekly user overlap} $U_{i,j,t}$ replicates the
procedure for \emph{user overlap} described above using the LSA model
computed over the entire study period, but the contributions from each
week.

\emph{Weekly topic overlap} $T_{i,j,t}$ between subreddits
is similar to weekly user overlap, but using token-frequency
inverse-document-frequency (TF-IDF) vectors instead of
author-frequency vectors.  Pointwise mutual information (PWMI) 
detects phrases for inclusion in the TF-IDF vectors.  Phrases of up to
four terms were used as it was the maximum computationally
tractable given the available resources. The number of possible phrases
grows exponentially with the phrase length.

To be included, a phrase must have a PWMI score of at least 3,
indicating that the phrase's probability is 20 times greater than the
product of the probabilities of each constitutive term. In addition, the
phrase count must be at least 3,500 within the 10\% sample and must
appear in at least 200 different subreddits. All single tokens appearing in at least 200 different subreddits also included.

Next, TF-IDF quantifies the ratio of the count of each token within
each subreddit divided by the maximum count for the subreddit and
by the log-count of the number of subreddits with the token.  LSA
reduces the dimensionality and sparsity of the resulting TF-IDF
while preserving subreddit similarities \cite{dumais_latent_2004}.
An HDBSCAN clustering procedure similar to that used to find
user-based clusters chose 1,000 dimensions for the topic overlap
LSA model.  The best topic-based clustering had 1,216 clusters
with 9,557 subreddits not assigned to any cluster.  The resulting
topic-based clusters are not used in subsequent analyses to follow
\citeauthor{teblunthuis_identifying_2022}'s use of user-based clusters
and because user-based clusters obtained a much greater silhouette
coefficient (0.53) than topic-based clusters
(0.34).  As with the author similarities, an LSA model
computed over the entire study period computes weekly topic overlap by
taking the cosine similarities of transformed weekly TF-IDF vectors
for each subreddit in the study period.  Both similarity measures
range from 0 (no overlap) to 1 (complete overlap).

\subsection{Panel Regression Models}


Fixed effect panel data estimators fit using ordinary least squares
test H2 and H3 \cite{wooldridge_econometric_2011}.  The
advantage of this method is that it is robust to confounding by
time-invariant unobserved variables.


The units of analysis in all of the regression models are dyadic
relationships between subreddits.  This dyadic structure violates the
assumption that observations are drawn independently from an identical
distribution. When a subreddit's topic changes, for instance,
resulting changes in its overlaps in all related subreddits will be
correlated.  Therefore, the analysis uses the ``cluster–robust variance estimation for dyadic
data'' technique developed by \citeauthor{aronow_clusterrobust_2015}
and implemented in the \texttt{dyadRobust} R
package\footnote{\url{https://rdrr.io/github/jbisbee1/dyadRobust/}} that 
provides consistent estimates of standard errors for dyadic data
with repeated measures \cite{aronow_clusterrobust_2015}. The Hypotheses
are tested via 95\% confidence intervals calculated by multiplying
the standard errors by 1.96. The dataset is sufficiently large to allow normal
distribution approximation.


\section{Results}

The analysis supports H1, that mutualism episodes are both more
frequent and longer lasting than competitive episodes.  However,
competitive episodes tended to be slightly stronger.  It also supports
H2, that competitive episodes lead to reductions in topic or user
overlaps between communities. However, it does not find evidence of
H3, that decreases in overlaps result in increasing mutualism.

\subsection{Mutualistic Interactions Are More Frequent and Longer Than Competitive Ones}

Although previous studies conceptualize mutualism and competition as static relationships, the present analysis finds that ecological interactions among overlapping online communities are bursty and relatively short-lived.
The average episode of competition or mutualism was only 2.13 weeks.

Mutualism episodes are slightly more frequent,
52.1\% of all episodes. They also last longer than
competition episodes. The average mutualism episode lasts
2.42 weeks and the average duration of
competition is 1.83.  The average
episode is weak mutualism with an average interaction strength of
0.04.

\begin{figure}

\includegraphics[width=\columnwidth]{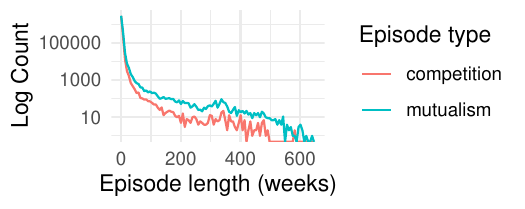} 

\caption{Frequency plot of the durations of competition and mutualism episodes. Mutualism tends to last longer than competition.  The y-axis is log-transformed. The axes are truncated to omit outliers for visibility.}
\label{fig:spell.lengths}
\end{figure}

Mutualism is more common at any given moment because it lasts longer.  Figure \ref{fig:spell.lengths} shows the distributions of the durations of competition and mutualism episodes in weeks. Most episodes are short: 80\% of competition episodes last 2  weeks or less, and  80\%  of mutualism episodes last 3 weeks or less. Among the longest episodes, most are mutualism. Of the top 5\% longest episodes, 100\% are mutualism.
The top 1\% of mutualism episodes last 13 weeks or more compared to the top 1\% of competition episodes which last at least 8 weeks.

\begin{figure}

\includegraphics[width=\columnwidth]{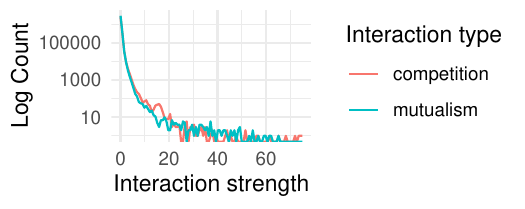} 

\caption{Frequency plot of average competition or mutualism strength for each episode. Competition and mutualism tend to have similar strengths. The y-axis is log-transformed.}
\label{fig:spell.strengths}

\end{figure}

\begin{figure}
\centering

\includegraphics[width=\columnwidth]{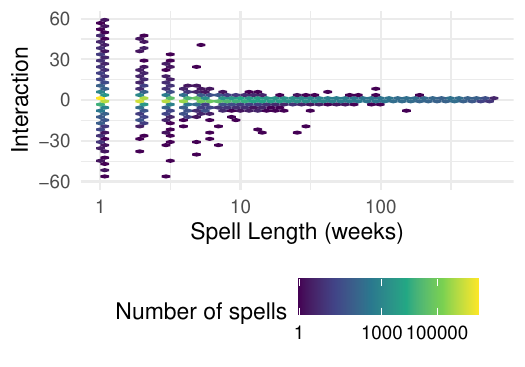} 

\caption{Episode lengths by the average competition and mutualism strength. Most episodes are short, but mutualism tends to last longer than competition. For clarity, the axes are truncated to remove extreme values.
\label{fig:spell.density}
}
\end{figure}

Competition and mutualism episodes are of similar strength on average, as Figures \ref{fig:spell.density} and \ref{fig:spell.strengths} visualize. The average competition strength is 0.2029 compared to 0.2039 for mutualism. 
As before, the extremes of the distribution are clarifying.
Most interactions are relatively weak: 50\% of mutualism episodes have an average $\mathcal{C}_{i,j}$ less than or equal to 0.12 and 50\% of competition episodes have an average $\mathcal{C}_{i,j}$ greater than or equal to -0.11. However, the strongest episodes are competition more often than mutualism. Of the 5\% strongest episodes, 51\%  are competition.
The 5\% strongest mutualism episodes have an average $\mathcal{C}_{i,j}$ greater than 0.62 and the top 5\% competition episodes have an average $\mathcal{C}_{i,j}$ less than -0.66. 

\subsection{Online Communities Increase Specialization in Relatively Competitive Conditions}

H2 proposed a positive (negative) relationship between mutualism
(competition) and future topic or user overlap because online
communities would specialize in response to competition. The evidence
from the panel regression models supports this hypothesis for both
types of overlap. Mutualism (competition) is positively associated
with increasing (decreasing) topic overlap ($B_1$=0.024;
CI=[0.0155, 0.0324]). The relationship
is relatively subtle. A one-unit increase in mutualism
$\mathcal{C}_{i,j}$ corresponds to an increase in term overlap by
0.024 standard deviations.

Similarly, online communities in mutualism (competition)
are likely to have subsequent increases (decreases) in user
overlap ($B_1$=0.02; CI=[0.0129,
0.0266]).  An increase of one unit 
mutualism strength $\mathcal{C}_{i,j}$ corresponds to an increase in user
overlap of 0.02 standard deviations; thus H2 is  supported.

\subsection{Reducing Overlaps Do Not Lead to Measurable Increases in Mutualism} 

Competition results in a decrease in niche overlap between
subreddits. Does this specialization, in turn, decrease competition
or increase mutualism, as H3 proposes? A negative (positive)
relationship between previous niche overlaps and subsequent mutualism
(competition) would support this hypothesis.  However, the panel
regression model instead has a positive relationship between previous
user overlaps and ecological interactions ($B_1$=0.06;
CI=[0.046, 0.074]), as well
as between previous topic overlaps and subsequent
mutualism ($B_2$=0.09; CI=[0.066,
0.113]), opposite in sign to that hypothesized. In sum,
the analysis finds no support for H3.

\section{Discussion}

Online community leaders and participants can build overlapping
communities to meet diverse and interconnected ne\-eds
\cite{teblunthuis_no_2022, zhu_impact_2014}.  This study investigates
how they do so with the a goal of providing designers with insights
into the relationships among overlapping communities. Prior
work treated competition and mutualism as static relationships
and found that mutualism was much more common than competition
\cite{teblunthuis_identifying_2022}. However, more flexible nonlinear time
series met\-hods reveal that these relationships are not static, but usually
take the form of episodic bursts. Findings from these models explain
that mutualism is more common than competition, not primarily because
mutualistic episodes have greater frequency or strength, but because
they last longer. Competitive episodes are slightly less common than
mutualistic ones and have similar strength on average; however mutualistic
interactions last over half a week longer on average.

How do online communities find their niches?  What gives rise
to frequent mutualism? Theories of competitive exclusion and
organizational ecology predict that high degrees of niche overlap lead
to competition \cite{hannan_organizational_1989}. By implication,
mutualistic online communities might be ``made'' via an adaptive
process if communities shift their niches in ways that decrease
competition and promote mutualism.  Alternatively, communities
``born'' in a competitive niche may struggle to sustain activity,
making those born in a mutualistic niche more likely to survive. This study
investigated these processes in a series of panel regression models
to test how ecological interactions might drive behavior patterns that
shift communities' niches to decrease user and topic overlaps. Although
this study finds evidence that competition tends to decrease niche
overlaps, it does not find that such changes subsequently reduce
competition. Thus, the adaptive ``made'' hypotheses seems inadequate
to explain the preponderance of mutualism over competition.


In fact, the results suggest that change tends to be negative
because it leads to increasing competition. These findings resonate
with previous research suggestive of structural inertia in online
communities \cite{halfaker_rise_2013, teblunthuis_revisiting_2018}.
Such inertia may limit the ability of communities to adapt to
competition.  In some cases, online communities have
significantly changed in response to the rise of competitors; this has
depended on major structural changes implemented by leadership and
has resulted in long-run participation declines \cite{hill_almost_2011}.
Moreover, the burstiness of mutualism and competition may itself be
a source of inertia. Even if a community perceives competition, any
urgency for responsive adaptation may be lost when the episode ends.

An alternative ecological explanation for frequent mutualism
is the variation, selection, and retention process
\cite{campbell_variation_1965, van_de_ven_explaining_1995}. Social
scientists in the mid-20th century used this framework to
explain sociocultural change in organizations and institutions, and it is still influential among organizational and cognitive
scientists today \cite{heyes_cognitive_2018,hannan_concepts_2019,
aldrich_organizations_2006, monge_evolution_2008}. A selection
process can produce groups of overlapping, specialized, and
frequently mutualistic online communities as follows: Online
community founders generate \emph{variation} in the form of newly
created online communities. Contributors \emph{select} communities
that provide the distinctive benefits that gratify them. Nascent
communities that successfully meet such members' needs will be
\emph{retained} as they bring together a critical mass of committed
members. Future work should develop and test such a model, perhaps
using agent-based models of how people join new online communities
\cite{foote_motivations_2023} or behavioral data on online
communities' early development \cite{tan_tracing_2018}.

\subsection{Design and Management Considerations}

Similar to ecologists caretaking natural environments, designers
and leaders of online communities may seek to intervene to manage
their ecosystems. They could introduce competitors of problematic
communities or develop mutualistic communities to complement existing
ones. Should they do so by building a new community or by adapting
an existing one?  Building new communities is notoriously difficult,
prone to failure, and seems slow compared to the timescales of
mutualism and competition \cite{schweik_internet_2012}; therefore,
adaptation might seem more practicable. However, this study's findings
suggest that communities face structural intertia and may struggle
to adapt mutualistically; still, solutions are conceivable. When
existing communities spawn new communities that have strong
connections with their parents, the new communities can grow faster
\cite{tan_tracing_2018}. Organizing a spin-off or offshoot 
community may thus be a viable strategy to fill needed roles in
online commuity ecosystems.  The ``parent'' community should create a
channel where members can collaborate to design the new community and
coordinate their efforts to build it.  Of course, doing so is only
possible through the efforts of interested members.

Alternatively, the study's finding that online communities do not
adapt to find mutualistic niches may reflect constraints that may
be surmounted by future designs. Community leaders will not attempt
a niche shift unless they see a connection between such change and
their community's goals. An \emph{ecological monitor} that detects
and displays competition episodes and opportunities for mutualism can
help them recognize such opportunities. Even if community leaders
recognize an opportunity, platforms such as Reddit offer limited
levers to execute change. They can alter their rules to differentiate
themselves from related communities; however, crafting and enforcing
rules requires time and effort, and change can be resisted or spark
migration \cite{kiene_identity_2024-1, davies_multi-scale_2021}.
Subtle changes to feeds or recommendations may nudge audiences and
contributors to gradually shift community niches. An implementation
challenge will be to ensure that such nudges do not violate users'
expectations or cause communities' niches to migrate beyond the bounds
of their topic and goals. Without such care, communities may resist
such designs.

\subsection{Threats to Validity}

Ecologists designed the regularized S-Map to infer competition and
mutualism in nonlinear systems, and it does so accurately over time up
to a constant \cite{cenci_non-parametric_2019}. However, this method
has limitations. Foremost, it does not account for uncertainty in the
estimation. In addition, it is possible that unobserved variables,
such as other communities not included in the models, may confound
these models. Finally, the method depends on a long time-series
of activity, which requires that the dataset exclude small and
short-lived subreddits whose ecological dynamics may differ from those
included. Future work should investigate other approaches to nonlinear
time series analysis that have different assumptions and can better
account for uncertainty \cite{kantz_nonlinear_2003}.

A second set of threats emerges from the panel regressions.  Although
fixed effect panel models are consistent even in the presence of
time-constant omitted variables \cite{wooldridge_econometric_2011},
time-varying unobserved variables, such as trends in topics' popularity, can potentially bias estimates. Therefore, readers should
not draw causal conclusions from these models. In general,
this study reveals the overall trends of competition, mutualism, and
overlaps on Reddit at a large scale. However, the wide view and analytical
approach obscure the particular mechanisms driving competition
and mutualism among these communities and when they are likely to
arise. Future fine-grained case studies and experiments should seek
to uncover such mechanisms.

Additional threats may result from weaknesses in the correspondence
between the study's measures and its scientific concepts.  The measure
of topic overlap depends on a relatively simple language model that
cannot fully capture the meaning of what people are saying. These
measures are relatively straightforward to understand; however, future work
may use more advanced language models to create more sensitive and
accurate measures of online community niches.  All of these measures are
based on observed activity within these communities; however, the analysis
includes communities that sometimes have little activity. Future work
should explore whether communities of different sizes tend to interact
similarly or differently and develop measures of online community
niches with greater precision in low-activity communities .

Furthermore, the inferences of competition and mutualism are based
on group size; however, group size is an incomplete definition of online
community success. Larger online communities are not necessarily
more successful; however, they provide different types of benefits than
small communities \cite{hwang_why_2021}. Similarly, as is common
in the literature on online community ecosystems, the measures of
membership and participation lump together a range of contributions
including both minor or repetitive contributions and intensive
efforts. Similarly, the measure of user overlap is based only on the 
number of contributions and does not account for qualitative
differences in participation. Future work should develop alternative
measures of online community success, participation, and membership.

Finally, this study analyzes the ecological dynamics within clusters of
overlapping communities on Reddit. The solution to the clustering
algorithm may not be unique, and analyses based on different
strategies for finding groups of overlapping subreddits may yield
different results.  Furthermore, although Reddit is among the largest
platforms for online communities, this study's findings may not
generalize to online communities on other platforms or during other time periods. Future research should investigate online community
ecosystems on other online community platforms.

\section{Conclusion}

Recent studies investigated relationships between overlapping
online communities using an ecological framework and found
widespread mutualistic relationships using time-series analysis
\cite{teblunthuis_identifying_2022}.  This study assumed that
ecological interactions between online communities are linear and
static; it also provided little explanation of how these mutualistic
interactions emerged. The present study uses nonlinear time-series
analysis methods from ecological studies of biological organisms to
reveal bursty ecological interactions that vary over time.  It finds
that mutualism is more common than competition because mutualistic
interactions are more common and, more importantly, longer lasting.
This study also tests a model for how online communities adapt to
avoid competition and increase mutualism. Although user and topic
overlaps decrease under relatively competitive conditions, evidence
does not support a theory that such specialization effectively reduces
competition or leads to mutualism. Future research should explain
the emergence of mutualistic interactions between overlapping online
communities in terms of how people organize and join communities early
in their development. Future designs may enable online community
ecosystem management by identifying open roles in the ecosystem and
then filling them by purposfully adapting existing communities or
organizing new ones.

\section{Ethical Statement}

This work's mode of inquiry is to analyze large-scale data from
social media that is publicly available. Some have concerns about
the ethics of such analysis, such as the possibility that the people
whose activities created the data we analyze would have been likely
to expect their activity would be so used. The author is sympathetic
to these serious concerns. That said, the nature of this analysis
exposes no individual to scrutiny and the reporting obscures even the
identities of the communities analyzed. As a result, the likely and
potential harms of this work to individuals are low.

The ultimate goal of this work is to improve online communities
through a scientific understanding of how they develop and relate
to each other. The negative consequences of this work if it is
used to improve online communities that cause harm are easy to
imagine. However, the author is optimistic in believing that the
benefits of this knowledge to socially beneficial communities will
ultimately outweigh the negative effects.

\section{Acknowledgments}

Text from a draft of this article was included as part of the         
author's PhD dissertation at the Department of Communication at       
the University of Washington. He thanks his committee: Professors    
Benjamin Mako Hill, Kirsten Foot, Aaron Shaw, David McDonald and      
Emma Spiro, for their generous support, wise advice and insightful    
comments.  Versions of this paper received helpful feedback at the    
International Communication Association's 2022 annual meeting and     
at the International Conference for Computational Social Science.     
Special thanks to Simone Cenci for advice on the regularized          
S-MAP. Thanks to the Community Data Science Collective for additional 
feedback and support.                                                 

Additional thanks to Jason Baumgartner and pushshift.io for providing
the Reddit data archive as well as to the Reddit users and communities
whose actions this study analyzes. Also, thanks to the peer reviewers
and program committee members whose insightful comments improved the
this article's quality. Any remaining errors and imperfections are
the author's.  This work was supported by NSF grants IIS-1908850 and
IIS-1910202 and GRFP \#2016220885 and was facilitated through the use
of the advanced computational infrastructure provided by the Hyak
supercomputer system at the University of Washington as well the Texas
Advanced Computing Center (TACC) at The University of Texas at Austin.

\bibliography{ms}

\end{document}